\documentclass[10pt, preprint]{aastex}
\usepackage{natbib}

\def\be{\begin{equation}}
\def\ee{\end{equation}}
\def\lesssim{\raisebox{-0.3ex}{\mbox{$\stackrel{<}{_\sim} \,$}}}
\def\gtrsim{\raisebox{-0.3ex}{\mbox{$\stackrel{>}{_\sim} \,$}}}

\begin{document}
\title{Curvature radiation in pulsar magnetospheric plasma}
\author{Janusz Gil\altaffilmark{1}, Yuri Lyubarsky\altaffilmark{2} \&  George I. Melikidze\altaffilmark{1,3}}
\altaffiltext{1}{Institute of Astronomy, University of Zielona G\'ora, Lubuska 2, 65-265,
Zielona G\'ora, Poland}
\altaffiltext{2}{Ben-Gurion University, Beer-Sheva, 84105 Israel}
\altaffiltext{3}{Abastumani Astrophysical Observatory, Al.Kazbegi ave. 2a, Tbilisi 380060, Georgia}

\begin{abstract}
We consider the curvature radiation of the point-like charge moving
relativistically along curved magnetic field lines through a
pulsar magnetospheric electron-positron plasma. We demonstrate
that the radiation power is largely suppressed as compared with
the vacuum case, but still at a considerable level, high enough to
explain the observed pulsar luminosities. The emitted radiation is polarized 
perpendicularly to the plane of the curved magnetic filed lines coincides with that of extraordinary waves,
which can freely escape from the magnetospheric plasma. Our
results strongly support the coherent curvature radiation
by the spark-associated solitons as a plausible mechanism of
pulsar radio emission.
\end{abstract}


\section{Introduction}

Although almost 35 years have passed since the discovery of
pulsars, the mechanism of their radio emission still remains
unknown. This is one of the most difficult problems of modern
astrophysics. Soon after discovery, curvature radiation was
suggested as a plausible mechanism for the observed pulsar radio
emission \citep{rc69, ko70, rs75}. In fact, curvature
radiation is the most natural and practically unavoidable emission
process in pulsar magnetosphere. Most of the pulsar models suggest
creation of dense electron-positron plasma near the polar cap.
These charged particles move relativistically with Lorentz factors
$\gamma\sim 10^2-10^3$ along dipolar magnetic field lines.
Therefore, they emit curvature radiation at the characteristic
frequencies $\nu_c\sim\gamma^3c/r_c$, where $r_c$ is the radius of
curvature of field lines, which falls into the observed pulsar
radio band if $r_c\sim 10^8 - 10^9$~cm. This is a typical value of
the radius of curvature of dipolar field lines at altitudes of
about $10^7 - 10^8$~cm, where the observed pulsar radio emission
is supposed to originate \citep[][ and references therein]{kg98}.

However, some serious problems are encountered while considering the
curvature radiation in pulsars. The first, and probably the most
important one, is related with coherency of the pulsar radiation.
It is well known that the incoherent sum of a single particle
curvature radiation is not enough to explain a very high
brightness temperature of pulsar radio emission. Therefore, one is
forced to postulate the existence of charged bunches containing at
least $10^{15}$ electron charges in a small volume that can
radiate the coherent curvature emission. However, it is not easy
to form such charge bunches \citep[see][ for review]{melr92}.
Moreover, even if a bunch can be formed, it is not automatic
that it will emit coherent radiation. For example, bunches formed
naturally by linear electrostatic waves \citep[e.g.][]{rs75}
cannot provide any emission \citep[hereafter MGP00, and references
therein]{mgp00}. The natural mechanism for the formation of
charged bunches was first proposed by \citet{karp75}, who argued
that the modulational instability in the turbulent plasma
generates charged solitons, provided that species of different
charge have different masses. Such charged solitons were observed
in the laboratory electron-ion plasma \citep{sagd79} and perhaps
even in the Earth ionosphere \citep{petv76}. In pulsar
magnetospheric plasma, distribution functions of electrons and
positrons are different, because plasma screens the electric field
induced by co-rotation \citep{sch74, cr77}. This causes the
effective relativistic masses of electrons and positrons to be
different, which can result in a net charge of solitons formed in
pulsar magnetosphere \citep[][MGP00]{mp80, mp84}. The net soliton
charge can also be induced by admixture of ions in the plasma flow
above the polar cap \citep{cr80, gmg03}. One should mention here
that to explain coherent radio emission we do not necessarily need
stable solitons but only large scale (as compared with Langmuir
wavelength) charge density fluctuations.

The second problem is related to the fact that the bunches
(solitons) are surrounded by the magnetized plasma, which strongly
affects the radiation process. The soliton size, which is
determined by the level of the turbulence, is evidently larger
than the wavelength of the Langmuir waves. A bunch is unable to
emit radiation with the wavelength shorter than its longitudinal
size. Therefore, the soliton should emit only at frequencies
$\omega$ below the frequency of the plasma waves. In the pulsar
frame of reference, the corresponding condition writes
$\omega<2\sqrt{\gamma}\omega_p$, where $\gamma$ is the Lorentz
factor of plasma motion, $\omega_p=(4\pi e^2n/m)^{1/2}$ is the
plasma frequency, $n$ is the total number density of electrons
and/or positrons, $e$ is and $m$ is the charge and the mass of
electron, respectively. Since at the expected emission altitudes
$\nu_p=\omega_p/2\pi \sim 1 $ GHz (see eq.~[3] in MGP00) and
$\gamma \sim 10^2$, the coherent radiation should be emitted at
frequencies $\nu \lesssim 20$ GHz, as observed in radio pulsars.

There are three waves propagating in the pulsar plasma: the
extraordinary wave $(\omega\approx kc)$ polarized perpendicularly
to the plane set by the ambient magnetic field and the wave vector
\mbox{\boldmath$k$}, and the two ordinary waves polarized in this
plane, that is, the superluminal wave $(\omega>kc)$ and subluminal
wave $(\omega<kc)$. Under the condition
$\omega<2\sqrt{\gamma}\omega_p$, emission of the superluminal wave
is heavily suppressed by the Razin effect \citep{razin60}. The
subluminal wave may, in principle, be emitted, but it cannot
escape from the plasma. In fact, it is ducted along the curved
magnetic field lines preserving direction of the wave vector and
eventually decays as a result of the Landau damping \citep{ba86}. The
extraordinary wave escapes freely but in the infinitely strong,
straight magnetic field it cannot be emitted because it does not
interact with the plasma particles. However, in the curved
magnetic field such wave can be emitted like in the vacuum case,
in which a significant fraction of the curvature emission is
polarized perpendicularly to the magnetic field line plane
\citep[e.g.][]{jack75, gs90}.

The aim of this paper is to calculate the curvature radiation
within the plasma under condition \be \varepsilon\equiv
\frac{\omega^2}{4\gamma\omega_p^2}\ll 1 . \label{cond} \ee We
consider emission from a point charge (modelling a small
soliton/bunch) moving circularly in the cylindrically symmetric
background. The magnetic field, which is assumed to be infinitely
strong, has circular field lines, and the electron-positron plasma
moves relativistically (with Lorentz factor $\gamma$) along these
field lines. We find that the point charge emits curvature
radiation in the extraordinary mode. The power of this radiation
is suppressed by a factor $\varepsilon$
(eq.~[\ref{cond}]), as compared with the vacuum case but still
remains rather high. It was found recently that the Vela pulsar
emits radio waves polarized predominantly in the direction
perpendicular to the plane of dipolar magnetic field lines
\citep{lai01}, strongly suggesting that these are escaping
extraordinary waves. Our results demonstrate that such radiation
can be generated within the pulsar magnetosphere by means of the
coherent curvature radiation. This radiation can also leave the
magnetospheric plasma and reach a distant observer.

MPG00 proposed the spark-associated soliton model for pulsar radio
emission, in which the charged solitons are formed by nonlinear
evolution of plasma waves triggered by a sparking discharge of
a high accelerating potential drop above the polar cap. These
relativistic solitons emit the coherent curvature radio emission
at altitudes of about 50 stellar radii in typical pulsars. The total luminosities
emitted by an ensamble of solitons are consistent with the
observed pulsar fluxes. The formation of charged solitons in
pulsar magnetosphere, as well as their coherent curvature
radiation, is well justified by MGP00. The only deficiency of this
model is that the curvature radiation was calculated in vacuum
approximation. We argue in this paper that the soliton coherent
curvature radiation is still a plausible mechanism of the observed
pulsar radio emission even when the presence of the ambient plasma
is taken into account.

\section{Wave equations}

\begin{figure}
\epsscale{0.4} \plotone{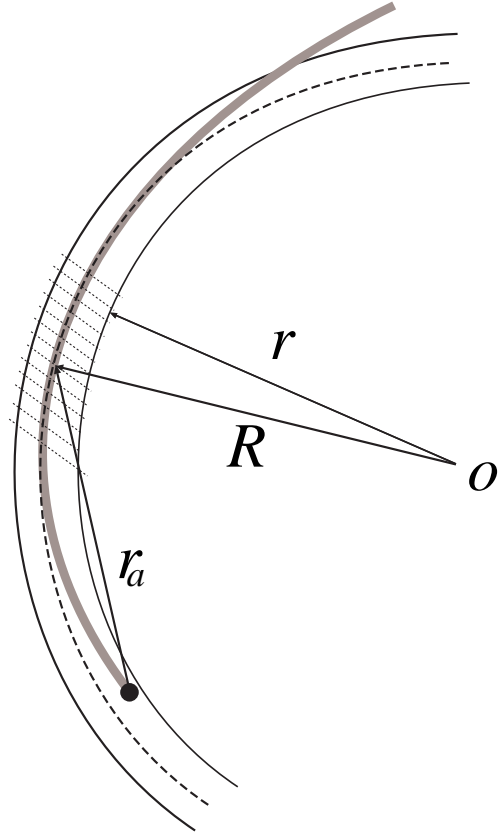} \caption{Schematic picture of
the radiation region (marked by dotted parallel lines) and the
system of cylindrical coordinates $(r, \theta, z)$, with $z$-axis
directed upright to the plane of the figure. The charged particle
moves along the dashed circle with radius $R$, which is locally
tangent to a dipolar magnetic field line marked by a thick shadowed
line. The thick black dot represents the pulsar and $r_a$ is radio
emission altitude.} \label{coordinate}
\end{figure}

We now examine the curvature radiation from the point-like
charge moving relativistically along a curved trajectory within the
electron-positron plasma. We find the radiation power by
calculating the work done on this charge by the excited
electromagnetic field \citep[e.g.][]{melr80}. It is convenient to
considered the curvature radiation in the cylindrical coordinates
$(r,\theta,z)$. The force lines of the infinitely strong magnetic
field are assumed to be concentric arcs with the radius of
curvature $r$ being much larger than the size of the considered
region of emission, marked by dotted parallel lines in Figure
\ref{coordinate}. The charged particle moves along the dashed
circle with radius R. The thick shadowed line represents the
actual magnetic field lines, which are locally modelled by
concentric arcs. The thick dot represents the pulsar and $r_a$ is
a radio emission altitude.

The wave equation for the electromagnetic vector potentials writes
\be
{\frac{1}{c^2}} \frac{\partial^2\mbox{\boldmath$A$}}{\partial
t^2}+\triangle
{\mbox{\boldmath$A$}}=-\frac{4\pi}{c}\mbox{\boldmath$J$},
\label{EqforA}
\ee
provided that vector and scalar potentials are
subject to the Lorentz gauge condition
\be
{\frac{1}{c}}
\frac{\partial\Psi}{\partial
t}+{{\bf\nabla}\cdot\mbox{\boldmath$A$}}=0. \label{EqforPhi}
\ee
The scalar potential $\Psi$ satisfies the wave equation as well,
provided that the total charge of the system is conserved. The
electric field of the wave is defined as
\be
\mbox{\boldmath$E$}=-{\frac{1}{c}}\frac{\partial
\mbox{\boldmath$A$}}{\partial t}-\nabla \Psi. \label{EqforE}
\ee
We use equations (\ref{EqforA}) and (\ref{EqforPhi}) as a complete
system of equations. Let us note that in the infinitely strong
magnetic field the particles are forced to move exclusively along
the field lines and, consequently, the current can be excited only
along the $\theta$-direction.

Making the Fourier transform \be
f(r,s,k_z,\omega)=\int\exp(i\omega t-is\theta-ik_zz)
f(r,\theta,z,t)dt d\theta dz, \label{Fourier} \ee where $f$
denotes $ \Psi$ and components of $\mbox{\boldmath$A$}$,
$\mbox{\boldmath$J$}$ and $\mbox{\boldmath$E$}$, we get the
following set of equations 
\be 
\Psi=s\frac{c}{\omega
r}A_{\theta}+\frac{k_z c}{\omega}A_z-i\frac{c}{\omega
r}\frac{\partial}{\partial r}rA_r; \label{Fpsi} 
\ee 
\be
\frac{\partial}{\partial r}\frac 1r\frac{\partial}{\partial
r}rA_r+\left(\frac{\omega^2}{c^2}-\frac{s^2}{r^2}-k^2_z
\right)A_r-2i\frac{s}{r^2}A_{\theta}=0; \label{FAr} 
\ee 
\be
\frac{\partial}{\partial r}\frac 1r\frac{\partial}{\partial
r}rA_{\theta}+\left(\frac{\omega^2}{c^2}- \frac{s^2}{r^2}-k^2_z
\right)A_{\theta}+2i\frac{s}{r^2}A_{r}=-\frac{4\pi}{c} J_{\theta};
\label{FAt} \ee \be \frac{\partial}{\partial r}\frac
1r\frac{\partial}{\partial
r}rA_z+\left(\frac{\omega^2}{c^2}-\frac{s^2}{r^2}-k^2_z
\right)A_z=0. \label{FAz} 
\ee 
One should remember that according
to the Landau prescription, $\omega$ is assumed to have a small
positive imaginary part. We do not write down it explicitly here
but will restore it in the appropriate place (see  bellow
eq.~[\ref{zerothAt}]).

Now we need to find the current density associated with the point
charge $J^{\rm part}$ and current density $J^{\rm plasma}$ excited
in plasma by the wave. If we assume that the emitting particle
rotates with the velocity $V$ (the corresponding Lorentz factor is
$\Gamma$) along a circle with the radius $R$, we can introduce
the corresponding current density as
$j_{\theta}=(QV/R)\delta(r-R)\delta(\theta-Vt/R)\delta(z)$, where
$Q$ is a charge of the emitting particle. Making the Fourier
transform we get \be J^{\rm part}_{\theta}=2\pi Q
\frac{V}{R}\delta(r-R)\,\delta\!\left(\omega-s\frac {V}{R}\right).
\label{FJpart} \ee In the infinitely strong magnetic field, only
$\theta$-component of the electric field of the emitted wave
excites the currents in the plasma. Assuming that the plasma is
cold and moves with the velocity $v$, we get \be J^{\rm
plasma}_{\theta}=i\frac{\omega_p^2}{4\pi}\frac{1}{\gamma^3}\frac{\omega}{\left(\omega-s\,\frac{v}{r}
\right)^2}E_{\theta}, \label{FJplasma} \ee (see details in
Appendix \ref{curinpl}). Here $\gamma=(1-v^2/c^2)^{-1/2}$ in the
plasma Lorentz factor. The Fourier component of the longitudinal
electric field (eq.~[\ref{EqforE}]) can be expressed via the
potentials in the form 
\be 
E_{\theta}=i\left(\frac{\omega}{c}
A_{\theta}-\frac{s}r\Psi\right). \label{FE} 
\ee 
Our objective is
to find such solution of equations (\ref{Fpsi}--\ref{FE}), which
remains finite as $r\to 0$ and comprises an outgoing wave as
$r\to\infty$.

\section{Short wavelength approximation}

Ultrarelativistic particles emit waves with the wavelength much
shorter than the curvature radius of their trajectory. Therefore,
we can solve the wave equations in the short wave approximation,
$\omega r/c \sim s \gg 1$. Let us start with the so-called WKB
approximation, in which $A\propto\exp(i\int k_rdr)$ and therefore
the solution splits into two waves \be
k_r^2=\frac{\omega^2}{c^2}-\frac{s^2}{r^2}-k_z^2 \label{wkbt} \ee
and \be
\omega^2-\left(\frac{s^2}{r^2}+k_z^2+k_r^2\right)c^2=\frac{\omega_p^2}{\gamma^3(\omega-s\frac
vr)^2}\left( \omega^2-\frac{s^2}{r^2}c^2\right). \label{wkblt} \ee
Equation (\ref{wkbt}) describes the extraordinary wave
($\omega=kc$) polarized perpendicularly to the plane set by the
local magnetic field and \mbox{\boldmath$k$} vector, while the
ordinary wave polarized in this plane is described by equation
(\ref{wkblt}). Under the condition expressed by equation
(\ref{cond}), equation (\ref{wkblt}) has a solution corresponding
to the nearly transverse wave with the dispersion law in the form
\be \omega^{2}=\left( \frac{s}{r}c\right)^{2}\left(
1-\frac{\left(k_{r}^{2}+k_{z}^{2}\right) c^{2}}{4\gamma^{2}\omega
_{p}^{2}}\right)\;\;{\rm if}\;\;\left( k_{r}^{2}+k_{z}^{2}\right)
c^{2}<<\omega _{p}^{2}. \label{wkbltsol} \ee The WKB approximation
is not valid close to the classical reflection radius $r_0$,
defined as \be \omega^2-\frac{s^2}{r_0^2}c^2-k^2_zc^2=0.
\label{rodef} \ee At $r\approx r_0$, the refraction indices
$N\equiv kc/\omega$ of extraordinary (eq.~[\ref{wkbt}]) and
ordinary (eq.~[\ref{wkbltsol}]) waves are very close to each other
($\omega=kc$), thus providing conditions for the wave coupling. In
this region $k_r$ goes to zero, thus forming a caustic zone.

Consequently, we should solve differential equations
(\ref{Fpsi}--\ref{FE}) for the electromagnetic potentials in
the region $r\approx r_0$ and then match the obtained solution
with the appropriate WKB solution (Eqs.~[\ref{wkbt}] or
[\ref{wkblt}]). Note that because the particle emits nearly along
the local magnetic field, the radius of the particle trajectory
$R$ should be also close to $r_0$. It is therefore convenient to
use the radial variable \be x\equiv \frac{r-r_0}{r_0}s^{2/3}.
\label{xdef} \ee In the region of interest $x\sim 1$ and
correspondingly $\vert r-r_0\vert\sim r_0 s^{-2/3}\ll r_0$. The
power series expansion in small $s^{-1/3}\ll1$ allows to reduce
equations (\ref{FAr}--\ref{FAz}) to the form \be
A_r^{\prime\prime}+2xA_r- 2is^{-1/3}A_{\theta}=0, \label{EArx} \ee
\be A_{\theta}^{\prime \prime }+2xA_{\theta}-
2is^{-{1}/{3}}A_{r}=D, \label{EAtx} \ee \be A_z^{\prime \prime
}+2xA_z=0, \label{EAzx} \ee (see Appendix \ref{waveeq} for
details), where
\begin{eqnarray}
D&\equiv&-4\pi r_0^2s^{-{4}/{3}} \left(J_{\theta}^{\rm{plasma}}+J_{\theta}^{\rm{part}}\right)\nonumber\\
&=&8\frac{\omega _{p}^{2}\gamma }{\omega ^{2}}\frac
{\left( 2\left( x+a\right) A_{\theta }+is^{{1}/{3}}A_{r}^{\prime }\right)}
{\left( 1+2s^{-{2}/{3}}\gamma ^{2}\left( a+x\right) \right)^{2}}
-8\pi^{2}VQs^{-{2}/{3}}\delta \left( x-X\right) \delta \left( \omega -s\frac{V}{R}\right),
\label{ED}
\end{eqnarray}
\be X=-a-\frac{1}{2}\frac{s^{{2}/{3}}}{\Gamma ^{2}}\;\;{\rm
and}\;\;\; a=\frac{1}{2}\frac{s^{{2}/{3}}k_{z}^{2}c^2}{\omega
^{2}}, \label{aXdef} \ee where prime denotes differentiation with
respect to $x$-coordinate and $X$ denotes $x$-coordinate of the
emitting particle (see Appendix \ref{waveeq}). We do not assume
any relationship between $A_r$ and $A_\theta$ and thus we retain
terms proportional to a small factor $s^{-1/3}$ in equations (\ref{EArx}) and
(\ref{EAtx}).

The gauge condition expressed by equation (\ref{EqforPhi}) still
permits a gauge transformation $\Psi\to\Psi-\partial\Phi/\partial
t$, ${\mbox{\boldmath$A$}}\to {\mbox{\boldmath$A$}}+\nabla\Phi$,
provided that $\Phi$ satisfies d'Alembert's equation, which
reduces to $\Phi^{\prime \prime }+2x\Phi=0$ if $\omega r\gg 1$ and
$\vert r-r_0\vert\ll r_0$. Therefore, one can choose $A_z=0$ (c.f.
eq.~[\ref{EAzx}]) and solve equations (\ref{EArx}) and
(\ref{EAtx}) together with equation (\ref{EqforPhi}) or equation (\ref{Fpsi}). 
The solution must comprise an evanescent wave at $x<X$ and an outgoing wave at
$x>X$.

The Green function for equations (\ref{EArx}) and (\ref{EAtx}) can 
be found from the solution  of the corresponding homogeneous
equations matched at the point $x=X$ according to the following
matching conditions
\be
A_{\theta}^{\prime}(X+0)-A^{\prime}_{\theta}(X-0)=-8\pi^2QVs^{-2/3}
\delta\left(\omega-s\frac VR\right)\;\;{\rm and}\;\;
A'_{r}(X+0)=A'_{r}(X-0), \label{mathcond1}
\ee
which can be
obtained straightforwardly integrating equations (\ref{EArx}) and
(\ref{EAtx}) between limits $X-0$ and $X+0$.

Let us consider the first term of $D$ (see eq.~[\ref{ED}], which
is just the right hand side of eq.~[\ref{EAtx}]). The nominator in
this term (which according to eq.~[\ref{AppB8E}] is proportional
to $E_\theta$) should be small, due to the condition expressed by
equation (\ref{cond}). Thus, in the zeroth approximation in
$\varepsilon$ (eq.~[\ref{cond}]) the following relationship holds
\be A_{\theta}=-i\frac{s^{1/3}}{2(x+a+i0)}A_r^{\prime}.
\label{zerothAt} 
\ee
Here we take into account the fact that
according to the Landau prescription $\omega$ and consequently $a$
have a small imaginary part. Substituting the obtained relation into
equation (\ref{EArx}) we get a closed equation for $A_r$. Let us note, that
equation (\ref{zerothAt}) means that $E_{\theta}=0$. Therefore in order to calculate 
the emission power we need the next approximation in $\varepsilon$ (see the next section).

Now it is convenient to introduce the dimensionless function
$u(x)$ such that 
\be 
A_r=\frac{8\pi^2
QV}{s}\frac{X+a}Xi\delta(\omega-s V/R)u, \label{zerothAr} 
\ee 
(see Appendix \ref{waveeq} for details), which satisfies the following
equation \be u^{\prime\prime}-\frac{u^{\prime}}{x+a+i0}+2xu=0.
\label{EU} \ee The matching conditions (eq.~[\ref{mathcond1}]) can
be now reduced to \be u(X+0)-u(X-0)=1\;{\rm and}\;
u^{\prime}(X+0)-u^{\prime}(X-0)=0. \label{mathcondu} \ee Equation
(\ref{EU}) has the Airy type asymptotics $u\to\exp(\pm
(2\sqrt{2}/3)ix^{3/2})$. Therefore, we should look for the
solution which satisfies the conditions
$u\to\exp(-(2\sqrt{2}/3)(-x)^{3/2})$ as $x\to -\infty$ and
$u\to\exp(i(2\sqrt{2}/3)x^{3/2})$ as $x\to \infty$. Then we can
find $A_r$ and $A_{\theta}$ from equations (\ref{zerothAt}) and
(\ref{zerothAr}), and match them to the WKB
solution (eq.~[\ref{wkbt}]) as $x\to\infty$. 

It is straightforward to calculate the electric field of the waves in the far zone using equations (\ref{EqforE}), 
(\ref{Fpsi}) and (\ref{zerothAt}) as
\be
\left(E_r, E_{\theta}, E_z\right)\propto\left(\frac{k_z c}{\omega},\; 0,\; -\frac{s^{1/3}}{\sqrt{2x}}
\right)\exp\left( i\frac{2\sqrt{2}}{3} x^{3/2}\right).
\label{polar}\ee
This solution is valid in the region $1\ll x \ll s^{2/3}$ (see eq. [\ref{xdef}]), where it is smoothly matched to the global WKB solution expressed by equation (\ref{wkbt}). One can see that the outgoing wave constitutes the extraordinary mode polarized predominantly in the direction perpendicular to the plane of the curved magnetic field lines.

Unfortunately
equation (\ref{EU}) cannot be solved by using any known special
functions, and has to be solved numerically.
For the numerical solution, it is convenient to reduce equation
(\ref{EU}) to the Rikkati type equation \be w^{\prime}+w^2-\frac
w{x+a+i0}+2x=0, \label{EW} \ee where $w=u^{\prime}/u$.  We need only the imaginary part of $u$,
because, as it is shown in the next section, the emissivity of the
particle is expressed via ${\rm Im} u(X)$ (see eq.~[\ref{EPgobal},\ref{firstE}]).
Using the matching conditions expressed by equation
(\ref{mathcondu}), we can express ${\rm Im }u(X)$ by the solution
of equation (\ref{EW}), that is 
\be 
{\rm Im}u(X)=\frac{w_{-}{\rm
Im}w_{+}} {(w_{-}-{\rm Re}w_{+})^2+({\rm Im}w_{+})^2}. \label{uw}
\ee
Starting
from a large negative $x$ for which $w=\sqrt{2}\sqrt{-x}$ (an
evanescent wave) and integrating numerically to $x=X$, we found a
solution $w_{-}\equiv w(X-i0)$. On the other hand starting from a
large positive $x$, for which $w=i\sqrt{2}\sqrt{x}$ (an outgoing
wave) and integrating to some $x=\epsilon\ll 1$, then integrating
(in the complex plane) along the semicircle
$x=\epsilon\exp{i\phi}$, and finally integrating further along the
real axis to the point $x=X$, we found another solution
$w_{+}\equiv w(X+i0)$.

\section{Emissivity of curvature radiation}

The emitted power can be found by calculating the work done by the
radiative electric field on the emitting particle \be P=-\int
dtd\mbox{\boldmath{$r$}}j^{\rm part}_{\theta}(t,{\bf
r})E_{\theta}(t,{\bf r}) \label{EPgobal} \ee
\citep[e.g.][]{melr80}. Note that $E_{\theta}=0$ in the zeroth approximation in
$\varepsilon$ (see eq.~[\ref{zerothAt}]), therefore
we should find the next approximation. It can be done easily
substituting into the left hand side of equation (\ref{EAtx}) the
values found in the zeroth approximation. We find that \be
E_{\theta }=2i\pi ^{2}s^{-1}\Gamma ^{4}\frac{k_{z}^{2}c}{\omega
_{p}^{2}}\frac{Q}{\gamma }\frac{\left( 1-\frac{\gamma ^{2}}{\Gamma
^{2}}\right) ^{2}}{\left( 1+\frac{k_{z}^{2}c^2}{\omega ^{2}}\Gamma
^{2}\right) }\delta \left( s-\omega \frac{R}{V}\right) u\left(
x\right). \label{firstE} \ee

\begin{figure}
\epsscale{0.5} \plotone{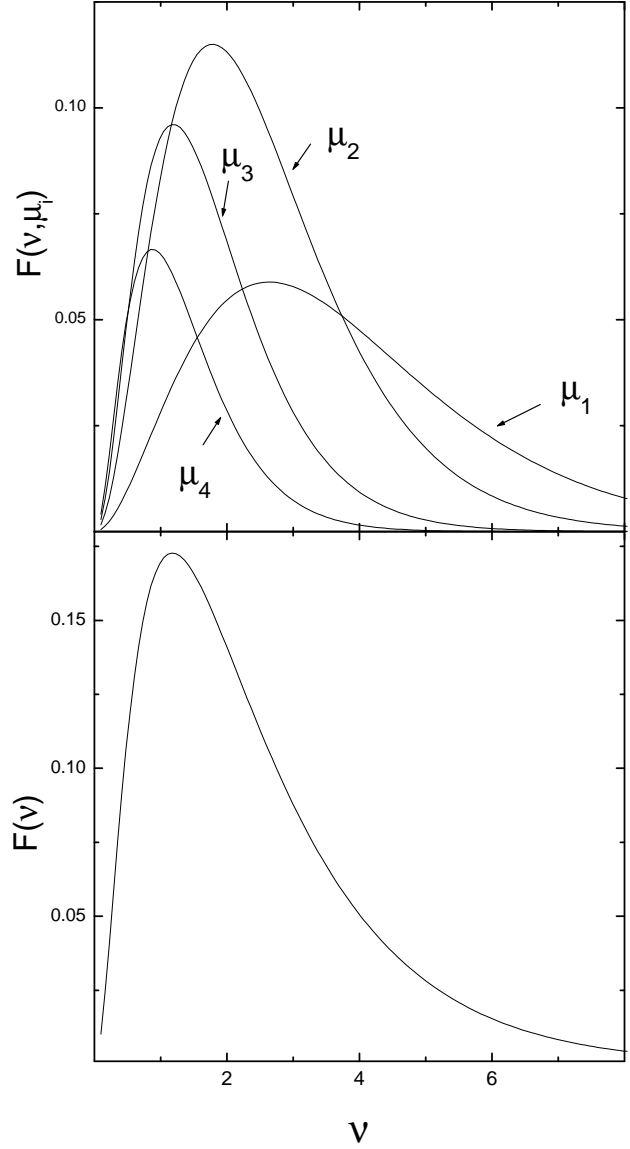} \caption{Distribution of the
radiation power emitted in all directions $F(\nu)=\int F(\nu,\mu)
d\mu$ (lower panel) and in various directions $F(\nu,\mu_i)$
(upper panel) over the dimensionless frequency
$\nu=\omega/\omega_c=\omega Rc^{-1}\Gamma^{-3}$, where
$\mu_1=0.5$, $\mu_2=1$, $\mu_3=1.5$ and $\mu_4=2$. }
\label{fvsfrequency}
\end{figure}

\begin{figure}
\epsscale{0.5} \plotone{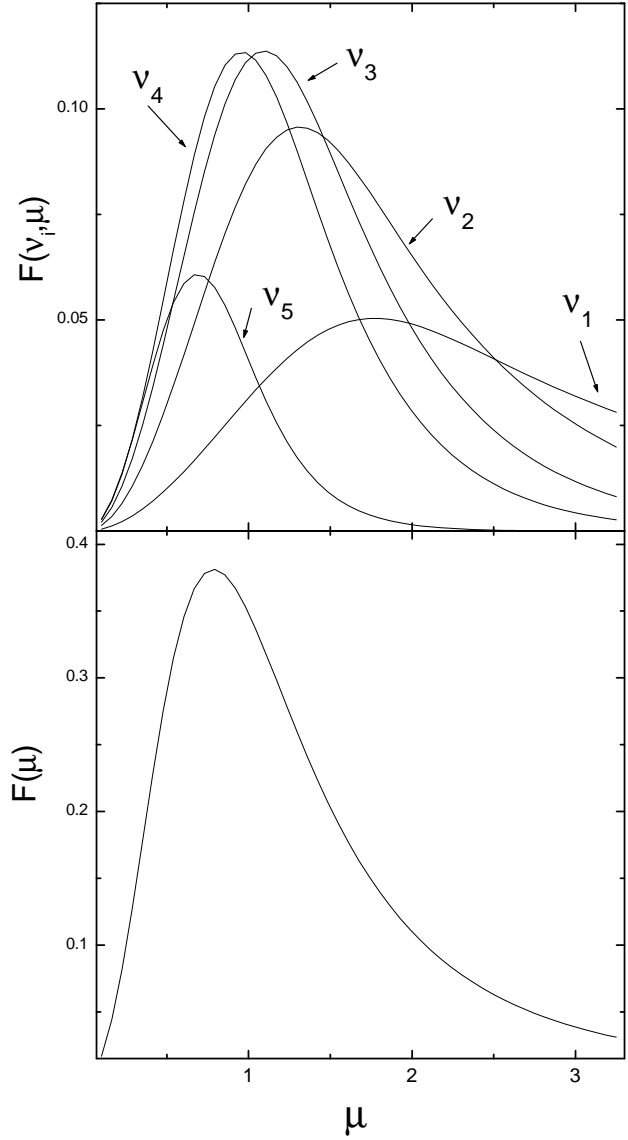} \caption{Distribution of the
radiation power emitted at all frequencies $F(\mu)=\int F(\nu,\mu)
d\nu$ (lower panel) and at various frequencies $F(\nu_i,\mu)$
(upper panel) over the dimensionless angle
$\mu=k_zc\Gamma/\omega$, where $\nu_1=0.5$, $\nu_2=1$,
$\nu_3=1.5$, $\nu_4=2$ and $\nu_5=4$.} \label{fvsangle}
\end{figure}

Now we can write the emissivity in the form 
\be 
\eta \equiv
\frac{d^{3}P}{dtd\nu d\mu }=\frac{c}{\pi }\Lambda
\frac{Q^{2}\Gamma ^{4}}{R^{2}}\left( 1-\frac{\gamma
^{2}}{\Gamma^{2}}\right) ^{2}F\left( \nu ,\mu \right), \label{eta}
\ee where \be F\left( \nu ,\mu \right) =\frac{\nu ^{2}\mu
^{2}}{\left( 1+\mu ^{2}\right) }{\rm{Im}}u\left( X\right),
\label{F} \ee and dimensionless frequency $\nu$ and angle $\mu$
are defined as \be \nu =\frac{\omega}{\omega_c}=\omega
\frac{R}{c\Gamma ^{3}}\;\;\;{\rm and}\;\;\;\mu
=\frac{k_z}{k}\Gamma=\frac{k_{z}c}{\omega }\Gamma. \label{numu}
\ee Let us note that a position $X$ of the emitting particle can
be expressed by means of $\nu$ and $\mu$ as $X=-0.5\nu
^{2/3}\left( 1+\mu^{2}\right)$. The parameter \be \Lambda=\frac
1{4\gamma}\left(\frac{c\Gamma^3}{\omega_pR}\right)^2\ll 1
\label{lambda} \ee describes the suppression of the emissivity of
the curvature radiation by the  surrounding plasma. Actually, this
is the ratio of the curvature radiation frequency to the plasma
waves frequency, thus it corresponds to the condition expressed by
equation (\ref{cond}) at the frequency $\omega=\omega_c\equiv
c\Gamma^3/R$ (see also discussion below eq.~[\ref{eL}]). As a
consistency check let us note that it follows from equation
(\ref{eta}) that $\eta = 0$ if $\Gamma=\gamma$ (no radiation from
uniform plasma motion), as expected. Function $F\left( \nu
,\mu \right)$ is plotted in Figures \ref{fvsfrequency} and
\ref{fvsangle} for various values of $\nu$ and $\mu$. Let us note
that $\mu = 1$ corresponds to the radiation angle $1/\Gamma$ from
the plane of the source motion and $\nu = 1$ corresponds to the
characteristic frequency of curvature radiation $\nu_c= c\Gamma^3/2\pi R$. 
The polarization of this radiation is that of the
extraordinary wave. The numerical values on the vertical axes in
Figs.~2 and 3 can be useful in estimating the emissivity of
curvature radiation of the point-like charge $Q$ moving
relativistically (Lorentz factor $\Gamma$) through the
relativistic plasma (Lorentz factor $\gamma$) in pulsar
magnetosphere. Note that the emissivity is zero if $\mu=0$, which means 
that although all the radiation is concentrated within the angle 
$\sim 1/\Gamma$ around the plane of the charge trajectory, there is no 
radiation exactly in this plane.

The total luminosity from the charge can be obtained
as $L_1=\int\eta d\mu d\nu$, where $\eta$ is the emissivity
described by equation~(\ref{eta}). One can find by numerical
integration that $\int F(\mu,\nu)d\mu d\nu=0.53$ and obtain 
\be
L_1\approx 0.1\Lambda c \frac{Q^2\Gamma^4}{R^{2}}\left(1-\frac{\gamma^2}{\Gamma^2}\right)^2. \label{eL1}
\ee 

\section{How can the particle emit radiation polarized perpendicularly to the plane of its trajectory?}

According to our results the escaping waves are polarized almost perpendicularly to 
the plane of the magnetic field curvature, whereas the wave itself propagates nearly in this plane.
Thus, this wave constitutes the extraordinary plasma mode. At first sight this conclusion seems counterintuitive.
It is generally believed that in the infinitely strong magnetic field the extraordinary wave cannot
be excited because the electric field of this wave is perpendicular to the external magnetic field and therefore it does not interact with the plasma. However, this statement is correct only in
the case of the straight magnetic field. It is the curvature of field lines which makes a difference. For example \citep{l93} and \citep{lm95} demonstrated that the extraordinary mode can be generated in the infinitely strong magnetic field provided that the field lines are twisted. In this paper we show that the extraordinary mode can be excited even if the curved field lines lie in the plane, as it is in the case of a dipolar pulsar magnetic field.

To illustrate that let us consider the vacuum case first. The curvature radiation is polarized predominantly in the plane of the particle trajectory, however about 15\% of the total power is emitted with the polarization perpendicular to this plane \citep[][p. 675]{jack75}. Of course, the radiation which is emitted exactly in the plane of the particle trajectory is completely polarized in this plane. However, the radiation emitted at the angle about $1/\Gamma$ to the plane of the particle trajectory ($k_z/k\sim 1/\Gamma$) is elliptically polarized.

In the case of the plasma in the infinitely strong magnetic field, a charged particle can emit only waves with a nonzero component of the electric field along the external magnetic field. According to the standard classification, these waves are referred to as the ordinary plasma modes. Strictly speaking the WKB approximation is violated in the radiation formation region, and therefore one cannot introduce the normal modes there. However, one can speak in terms of the normal modes  heuristically. At the condition (\ref{cond}) the particle emits subluminal, quasi-transverse Alfven waves (the emission of the superluminal ordinary mode is suppressed by the Razin effect). The Alfven wave polarized in the plane of the curved external magnetic field does not propagate outwards, because it is ducted along the curved magnetic field and eventually decays by the Landau damping \citep{ba86}. However, just like in the vacuum case, there are waves with $k_z/k\sim 1/\Gamma$ and they are polarized perpendicularly to the plane of the external magnetic field line. The polarization plane of such a wave would rotate by the angle about unity as it propagates the distance $l\sim R k_z/k\sim R/\Gamma$, if the adiabatical walking condition
\be
(N-1)kl \gg 1
\label{condadia}
\ee
was satisfied \citep{cr79}. Here $N$ is the refraction index of the wave, which in our case can be estimated as 
\be
N-1\sim \frac{1}{2\Gamma^2}+\frac{1}{2}\left(\frac{k_z}{k}\right)^2\sim \frac{1}{\Gamma^2},
\ee 
\citep[see e.g.][or eq.(15)]{ba86}.
Taking into account that $k\sim \Gamma^3/R$, one gets straightforwardly $(N-1)kl\sim 1$. So, the adiabatical walking condition (\ref{condadia}) is not satisfied and therefore the wave escapes from the plasma retaining the initial polarization in the direction perpendicular to the magnetic field lines plane. In the far zone this wave should be classified as the extraordinary wave. Thus the extraordinary mode can be emitted in the curved infinitely strong magnetic field via the linear coupling of the normal modes in the radiation formation region.

It is interesting to compare the expression for the luminosity of the curvature radiation in plasma (Eq.[\ref{eL1}]) with the power of single-particle curvature radiation in vacuum
$p=(2/3)cQ^2\Gamma^4R^{-2}$ \citep{jack75}. As one can see, equation
(\ref{eL1}) has a quite intuitive form. The coefficient
$0.1\approx(2/3)(1/7)$ is related to the fact that this luminosity corresponds to the "perpendicularly" polarized  radiation, which contains only about 1/7 of the full power of curvature radiation \citep{jack75}.
Thus, the term $\Lambda\left(1-{\gamma^2}/{\Gamma^2}\right)^2$ in equation~(\ref{eL1}) represents the influence of
the plasma, while the factor $0.1$ represents the weaker 
extraordinary mode of curvature radiation.

\section{Implications for pulsars and discussion}

The two-stream instability induced by a sparking discharge of the
high potential drop above the polar cap  easily excites the strong
Langmuir turbulence at altitudes $10^{7} - 10^{8}$ cm
\citep{usov87, am98}, where the pulsar radiation is supposed to
originate \citep[e.g.][]{kg98}. As shown by MGP00 this instability
can lead to the formation of charged relativistic solitons in
pulsar magnetospheric plasma, which may play a role of point-like
charged bunches that can emit coherent curvature radiation. The
size of the solitons is about the characteristic instability
length-scale, which can be estimated as 
\be
d{^\prime}\sim\left(\frac{\omega{^\prime}_p}c\right)^{-1}
\left(\frac{W{^\prime}}{mc^2n{^\prime}}\right)^{-1/2},
\label{size1} \ee 
where $W$ is the energy density of the Langmuir
waves and the characteristic plasma ``temperature'' (the kinetic
energy spread) is about $mc^2$. All primed values refer to the
proper plasma frame of reference, in which the transverse size of
the soliton should be about the same as the longitudinal size (by
the causality argument). Taking into account that in the pulsar
frame the longitudinal size decreases $\gamma$ times whereas the
transverse size remains the same, one can assume that our bunches
have a form of pancakes. The soliton emits at wavelengths about
its longitudinal size $\lambda\sim d=d'/\gamma$, therefore its
transverse size is about $\gamma d$.  Note that emission from
two points separated by the transverse distance $d_{\bot}$ is added
coherently only within the angle $\sim\lambda/d_{\bot}$. Because the
relativistic emission is concentrated within the angle $\sim
1/\gamma$, one can see that the transverse size of a coherently
radiating bunch cannot exceed $\sim\gamma d$.

The solitons in a real turbulence do not survive for a long time.
The stronger the turbulence, the shorter the soliton life time. The
soliton emits as a single entity if it survives for a radiation
formation time, which is defined as a time during which the phase
of the emitted waves changes by about $2\pi$ in the emitter frame
of reference. This time is about $\lambda/(c\Gamma)\sim
d/(c\Gamma)$ in the soliton frame. Because the soliton cannot
decay for a time shorter than it takes for the light to propagate across
the soliton itself, the above condition does not place any
restrictions on the proposed emission mechanism. What we need in
fact are not stable solitons, but only long wavelength
fluctuations, whose existence is guaranteed by the modulation
instability criterium (MPG00, eqs. [A20] and [A23]).

The coherency condition requires that the radiated waveleght is much larger 
than the longitudinal size of the emitting soliton. This is equivalent to the 
condition expressed by equation (\ref{cond}). Adopting $\omega_{c}={c\Gamma ^{3}}/{R}$ 
and $\omega _{p}=\left( {4\pi e^{2}n}/{m}\right)^{1/2}$, where  $n=\kappa n_{{\rm GJ}}=1.4\times 10^{11}\kappa r_{6}^{-3}\left( \dot{P}_{-15}/P\right)^{1/2}$ cm$^{-3}$ is the plasma number density 
at the altitude $r=r_{6}\times 10^{6}$ cm, $n_{\rm GJ}$ is the \citet{gj69} number density, $R\sim 10r$ is the radius of curvature of dipolar field lines, $\Gamma =\Gamma_2 100$,  $\gamma =\gamma_2 100$ and $\kappa$ is the \citet{stu71} multiplication factor, we obtain 
\be
\varepsilon =10^{-4}\left( \frac{r_{6}}{\kappa }\right) \left( \frac{\Gamma_{2}^{6}}{\gamma _{2}}\right)\left( \frac{P}{\dot{P}_{-15}}\right)^{0.5},
\label{epscond}
\ee
which should be much less than unity (say at least $0.1$), where $P$ is the pulsar period in seconds and
$\dot{P}_{-15}=\dot{P}/10^{-15}$ is the period derivative in units
of $10^{-15}$~s/s. The actual value of $\kappa $ is somewhat 
uncertain, with estimates ranging from $1$ to $10^{4}.$ The lower values of $\kappa <100$ correspond 
to the stationary acceleration models \citep[e.g.][]{ha01}, while in the non-stationary sparking scenario 
being of interest here, one can expect larger values of $\kappa \approx 10^{2}-10^{4}$ \citep{rs75, melr00}. However we will 
estimate $\varepsilon $ as a function of $\kappa $ treated as a free parameter. One can see from the above equation 
that for emission altitudes $10 < r_{6} < 100$ \citep{kg98} $\varepsilon \lesssim 0.1$ if $\kappa \gtrsim 100$ (for $P=\dot{P}_{-15}=1$, $\gamma_2=2 $ and $\Gamma_2=4$). Thus, the basic condition (eq.[\ref{cond}]) for the spark associated soliton curvature radiation is certainly satisfied. Now let us consider the energetics of relevant processes.

\subsection{Radiation luminosity of the turbulent plasma}

Let us consider the emitting region within a tube of open dipolar
field lines, whose radial extent $\Delta r\sim r_a$ (see Fig.
\ref{coordinate}) and the cross-section surface area is $S$. The
total number of solitons can be estimated as $N\sim Sr_a/(2{\cal
V})$, where ${\cal V}$ is the soliton volume, and we assumed that
$\lambda \gg c\pi/{\sqrt \gamma}\omega_p$, thus $\varepsilon\sim \Lambda\ll 1$ 
(eqs. [\ref{cond}] and [\ref{epscond}]). 
Assuming reasonably that $\Gamma\sim 2\gamma$ (see MPG00) the total luminosity of ${\cal N}$ solitons  may be estimated as 
\be
L\approx 0.1\Lambda c Q^2\Gamma^4R^{-2}{\cal N} , \label{eL}
\ee where $Q$
is the point charge moving relativistically (with Lorentz factor
$\Gamma$) along a circle with radius $R$ and
$\Lambda=\omega^2_c/4\omega_p^2\gamma$ is the
suppression factor defined by equation~(\ref{lambda}).
The charge of the soliton is $Q=e\delta n{\cal V}$, where $\delta n$ is the
difference between the densities of positrons and electrons within
the soliton.
Since the charge 
density within the soliton is determined by the Langmuir wave energy density $W$ (e.g. MGP00), then we have
\be
\frac{\delta n}{n}\sim\left(\frac{W}{nmc^2\gamma}\right)^2
.\label{deltaen}
\ee
Noting that
$Q^{2}{\cal N}=(1/2)e^{2}\left( \delta n/n\right)^{2}n^{2}r_{a}{\cal V}S=\left( 1/8\pi \right) \left( 1/\gamma c^{3}\right) \left( \delta n/n\right) ^{2}\omega _{p}^{2}r_{a}{\cal V}L_{0}$,
where $L_0\equiv nmc^3\gamma S$ is the total kinematic power of
the plasma flow (section \ref{pkl}), and remembering that the soliton volume is estimated from equation (\ref{size1}), the total soliton
luminosity can be written in the form 
\be
L\sim 2.5\times \varepsilon^{1.5}\left( \frac{\Gamma _{2}^{3}}{\gamma _{2}^{2}}\right) \left( \frac{W}{\gamma mnc^{2}}\right)^{2.5}L_{0}.
\label{lumin2}
\ee
Here we assumed that $r_a/R\sim 0.1$ for dipolar field lines
at altitudes $r_a\sim(10^7 - 10^8)$~cm. 
This luminosity can be compared with the observed pulsar radio luminosity
(sections \ref{prl} and \ref{scd}).

\subsection{Pulsar kinematic luminosity}\label{pkl}

Since the kinematic flux is conserved along the tube of open
dipolar field lines, we can calculate $L_0=nmc^3\gamma S$ near the
polar cap, where $r_6$ is about $1$ and $S=\pi r_p^2\sim 3\times 10^8P^{-1}{\rm cm}^2$ 
is the polar cap surface area. Thus, the kinematic luminosity 
\be 
L_0\approx
10^{24}\gamma \kappa \dot{P}_{-15}^{1/2}P^{-3/2}\,\,{\rm erg~s^{-1}}.
\label{L0} 
\ee 
Since the product
$0.05\lesssim\dot{P}_{-15}^{1/2}P^{-3/2}\lesssim 10^2$, with a
typical value being about unity and $\gamma\sim 100$ 
we have a typical value ${\cal L}_0 \sim\kappa 10^{26}$~erg s$^{-1}$, where $\kappa> 100$,

\subsection{Observed pulsar radio luminosity}\label{prl}

The pulsar radio luminosities $L_R$ can be obtained from measured
fluxes and estimated pulsar distances. If $l=log(S_{400}D^2)$,
where $S_{400}$ is the mean flux density at 400 MHz given in mJy
and $D$ is the pulsar distance in kpc, then \be L_R=3.5\times
10^{25+l}~{\rm erg\ s}^{-1}. \label{LR} \ee According to Figure~9
in \citet[][ Pulsar Catalogue]{tml93}, $1\lesssim l\lesssim 3$,
with median value $<l>\sim 2$. Thus, the pulsar radio luminosities
$3.5\times 10^{26}\lesssim L_R\lesssim 3.5\times 10^{28}$~(erg
s$^{-1}$), with median value ${\cal L}_R\simeq 3.5\times
10^{27}$~erg $s^{-1}$.
\subsection{Comparison with observed luminosities}\label{scd}
Now, using equations (\ref{epscond}) and (\ref{L0}) we can rewrite equation (\ref{lumin2}) in the form
\be
L=2.5\times 10^{20}\left( {P^{-0.75}}{P_{-15}^{-0.25}}\right)
\left( \frac{r_{6}^{1.5}}{\kappa^{0.5}}\right)\left( \frac{\Gamma _{2}^{12}}{\gamma _{2}^{3.5}}\right) 
\left( \frac{W}{\gamma m_{e}nc^{2}}\right)^{2.5}
\label{L}.
\ee
Assuming  $\gamma_2\sim 2$, $\Gamma_2\sim 4$ and $\kappa\sim 100$ we can 
estimate the luminosity as $L\sim 5\times 10^{27}$ erg s$^{-1}$ 
(close to the median value ${\cal L}_R$), provided that $\left({W}/{\gamma m_{e}nc^{2}}\right)\sim 0.5$.
This requirement does not seem too excessive. In fact, the two-stream
instability due to overlapping of adjacent plasma clouds
associated with successive sparks \citep{usov87, am98} should
provide a high enough level of turbulence, because energies of
the plasma and the beam are of the same order.
Thus, one can expect that about 50\% of the beam energy will
be transferred to the plasma waves. Let us note that the luminosity is quite sensitive
to estimations of the soliton volume. The longitudinal size of the soliton is defined by 
the Langmuir turbulence (eq.[\ref{size1}]), but the cross-section may can be estimated using
coherency conditions according to the radiated wavelength. Then the luminosity can be higher
by at least one order of magnitude.
 Moreover, the term involving $\gamma_2$ and 
$\Gamma_2$ can drastically increase 
the value of $L$ under minor changes of Lorentz factors.
Therefore, the model of
curvature radiation of the spark-associated solitons developed by
MGP00 in vacuum approximation is still a plausible explanation of
pulsar coherent radiation, once the influence of the ambient
plasma is taken into account. The power of curvature radiation is
suppressed by plasma but not drastically with respect to the
vacuum case. The polarization of curvature radiation emitted in
plasma is that of the extraordinary mode, so it can escape from
the pulsar magnetosphere. 

Interestingly, \citet{lai01} argued recently that
the polarization direction of radio waves received from the Vela pulsar is perpendicular to the planes of dipolar magnetic field lines. It is instructive to follow their argument, which strongly implies that 
the observed radiation from this pulsar represents the extraordinary plasma mode. In fact,
\citet{lai01} were able to demonstrate that in the fiducial phase corresponding to the fiducial plane
(containing the rotation and the magnetic axes as well as the line-of-sight) the radiation is polarized 
perpendicularly to the plane of the dipolar magnetic field lines. This argument can be extended to every phase
within the pulse window, since the position angle swing in this pulsar is known to satisfy perfectly 
the geometrical rotation vector model \citep{rc69}. This means that in the Vela pulsar the polarization of
observed radio emission is consistent with the curvature radiation originating
in pulsar magnetospheric plasma. 

\acknowledgments This paper is supported in part by the Grant
2~P03D~008~19 of the Polish State Committee for Scientific
Research. We thank Don Melrose for his hospitality during our 
staying at the School of Physics (Sydney University),
where the final version of the paper was completed. 
We thank E. Gil and U. Maciejewska for technical help.

\appendix

\section{Current in the plasma\label{curinpl}}

In order to calculate the current density in the plasma let us
start with the equation of motion which in the relativistic case
writes \be m\gamma^{3}\frac{dv}{dt}=eE_{\theta }. \label{eqmotion}
\ee In the Fourier components we have
$$
\frac{dv}{dt}=-i\left( \omega -s \frac{d\theta}{dt}\right)v.
$$
Since $d\theta/dt=v/r$, thus the equation of motion writes

\be
v=\frac{i}{\gamma^{3}}\frac{e}{m}\frac{E_{\theta }}{\left( \omega -\frac{s}{r}v\right)}.
\label{eqmotFur}
\ee
The electric field is expressed by equation (\ref{FE}). Let us use the continuity equation in the form
$$
\frac{\partial n}{\partial t}+\nabla\cdot(n\mbox{\boldmath$v$})=0.
$$
Assuming that the perturbation of the density can be expressed as $n\rightarrow n+\delta n$, then for the corresponding Fourier components we can write $\partial n/\partial t = -i\omega \delta n$ and $\nabla\cdot(n\mbox{\boldmath$v$})=\mbox{\boldmath$v$}\nabla n + n \nabla\cdot \mbox{\boldmath$v$}=iv(s/r)\delta n+iv(s/r)n$. Then using equation (\ref{eqmotFur}) we obtain
\be
\delta n=\frac{i}{\gamma ^{3}}\frac{q_{e}n}{m}\frac{s}{r}\frac{E_{\theta }}{\left( \omega -\frac{s}{r}v\right)^{2}}.
\label{deltanF}
\ee
Assuming that the unperturbed charge density equals to zero, we can express the charge density perturbation as $\rho=2e\delta n$. Now we can find the current density from the following equation
$$
\frac{\partial \rho }{\partial t}+\nabla\mbox{\boldmath$J$}=0,
$$
which in the Fourier transforms writes
\be
-i\omega \rho +i\frac{s}{r}J_{\theta}=0.
\label{FCPA}
\ee
Substituting equation (\ref{deltanF}) into (\ref{FCPA}) we arrive at equation (\ref{FJplasma}).

\section{Wave equation in short wave approximation\label{waveeq}}

We are looking for the solution of equations (\ref{Fpsi} -- \ref{FAz}) in an narrow region
$r-r_0\ll r_0$, therefore we can substitute $r$ with $r_0$ into coefficients,
which vary slightly in this region. Then we can expand the coefficients which
vary significantly to the first order in $r-r_0$, those are the coefficients in the curly
brackets in equations (\ref{FAr} -- \ref{FAz}), which become zero at $r=r_0$ and the coefficient in the
denominator of equation (\ref{FJplasma}), which is close to zero at $v\rightarrow c$ and $k_z\ll \omega/c$.
Using the approximation $s\gg1$, $k_zc\ll\omega$ and $\gamma\gg1$
from definitions expressed by equations (\ref{rodef}) and
(\ref{xdef}) we get \be \omega =\frac{s}{r_{0}}c\left(
1+\frac{1}{2}\frac{r_{0}^{2}}{s^{2}}k_{z}^{2}\right) \label{AppB1}
\ee and \be \frac{1}{r}=\frac{1}{r_{0}}\left( 1-s^{-2/3}x\right).
\label{AppB2} \ee Then using a definition \be
a=\frac{1}{2}s^{2/3}\frac{k^{2}_{z}c}{\omega ^{2}} \; ,
\label{AppB4a} \ee it is straightforward to obtain the following
formulae 
\be
\frac{\omega^2}{c^2}-\frac{s^2}{r^2}-k^2_z=2\frac{s^{4/3}}{r_0^2}x
\label{AppB3} \ee and \be \left( \omega -\frac{s}{r}v\right)
=\frac{1}{2\gamma ^{2}}\frac{sc}{r_{0}}\left( 1+\left(
\frac{\gamma k_z c}{\omega
}\right)^{2}+2s^{-{2}/{3}}\gamma^{2}x\right)=\frac{1}{2\gamma ^{2}}\frac{sc}{r_{0}}\left(
1+2s^{-{2}/{3}}\gamma ^{2}\left( a+x\right) \right),
\label{AppB4} 
\ee

We also need an evaluation of expressions for the currents (see
eqs. [\ref{FJpart}] and [\ref{FJplasma}]). The argument of the
first $\delta$-function in equation (\ref{FJpart}) can be
evaluated using the condition $\omega R=sV$, which follows from
the argument of the second $\delta$-function. Therefore, we get
\be \delta \left( r-R\right) =\frac{s^{2/3}}{r_{0}}\delta \left(
x-X\right), \label{AppB5} \ee where \be X=-\frac{s^{2/3}}{2}\left(
\frac{k_{z}^{2}c^2}{\omega ^{2}}+\frac{1}{\Gamma ^{2}}\right)
\label{AppB6X} \ee describes the position of the emitting
particle. Using equation (\ref{AppB5}), we can express the current
density excited by the emitting particle (eq.~[\ref{FJpart}]) as
\be 4\pi r_{0}^{2}s^{-{4}/{3}}J_{\theta }^{\rm{part}}=8\pi
^{2}s^{-{2}/{3}}VQ\delta \left( x-X\right) \delta \left( \omega
-s\frac{V}{R}\right). \label{AppB7Jp} \ee Using equation
(\ref{Fpsi}) we can rewrite the expression for the electric field
(eq.~[\ref{FE}]) in the form \be E_{\theta }=i\frac{c}{\omega
}\left( 2\frac{s^{{4}/{3}}}{r_{0}^{2}}x+k_{z}^{2}\right) A_{\theta
}-\frac{c}{\omega }\frac{s^{\frac{5}{3}}}{r_{0}^{2}}A_{r}^{\prime
}, \label{AppB8E} \ee and with the use of equations (\ref{AppB4})
and (\ref{AppB8E}) the expression for the plasma current density
(eq.~[\ref{FJplasma}]) can be written as \be -4\pi
r_{0}^{2}s^{-{4}/{3}}J_{2}^{\rm{plasma}}=8\frac{\omega
_{p}^{2}\gamma }{\omega ^{2}}\left( 1+2s^{-{2}/{3}}\gamma
^{2}\left( a+x\right) \right)^{-2}\left( 2\left( x+a\right)
A_{\theta }+is^{{1}/{3}}A_{r}^{\prime }\right).
\label{AppB9Jplasma} \ee Now using equations (\ref{AppB7Jp}) and
(\ref{AppB9Jplasma}) we obtain the right hand side of equation
(\ref{EAtx}) as it is defined by equation (\ref{ED}).

Now we need to find the matching condition for $A_{r}$. Using
equation (\ref{zerothAt}) we have \be A_{\theta }^{\prime
}=-\frac{i}{2}\frac{s^{{1}/{3}}}{\left( x+a+i0\right)
}A_{r}^{\prime \prime }-\frac{1}{\left( x+a+i0\right) }A_{\theta
}, \label{AppBAtder} \ee and getting $A_{r}^{\prime \prime }$ from
equation (\ref{EArx}) and substituting it into equation
(\ref{AppBAtder}) we get \be
A_{r}=-is^{-{1}/{3}}\frac{x+a}{x}A_{\theta }^{\prime }.
\label{AppBArdef} \ee Now using the matching condition for
$A_{\theta }^{\prime }$ (eq.~[\ref{mathcond1}]) we obtain the
following matching condition for $A_{r}$ \be A_{r}\left(
X+0\right) -A_{r}\left( X-0\right) =i8\pi ^{2}V Q s^{-1}\delta
\left( \omega -s\frac{V}{R}\right) \frac{X+a}{X}.
\label{mathcond2} \ee With the help of the above equation, the
matching condition expressed by equation (\ref{mathcondu}) follows
from definition expressed by equation (\ref{zerothAr}).

\section{Emissivity \label{emis}}

The reverse Fourier transform writes (c.f. eq.~[\ref{Fourier}])
\be f\left( \theta ,z,t\right) =-\frac{1}{\left( 2\pi \right)
^{3}}\int\limits_{-\infty }^{\infty }\int\limits_{-\infty
}^{\infty }\int\limits_{-\infty }^{\infty }f\left( s,k_z,\omega
\right) \exp \left( -i\left( \omega t-k_zz-s\theta \right) \right)
d\omega dk_zds \label{revfur}. \ee Substituting the Fourier
transforms of the current density and the electric field to
equation (\ref{EPgobal}) we obtain
\begin{eqnarray}
P&=&-\frac{1}{\left( 2\pi \right) ^{6}}\int\limits_{-\infty }^{\infty }rdrd\theta dzdtJ_{\theta}^{\rm{part}}\left( \omega ^{\prime },r,s^{\prime },k_{z}^{\prime }\right) E_{\theta}^{\ast }\left( \omega ,r,s ,k_{z}\right)\nonumber\\ &\times& \exp \left( i\left( \omega -\omega ^{\prime }\right) t-i\left( k_{z}-k_{z}^{\prime }\right) z-i\left( s-s^{\prime }\right) \theta \right) d\omega dk_{z}dsd\omega ^{\prime }dk_{z}^{\prime }ds^{\prime },
\label{P1}
\end{eqnarray}
where $E^*_\theta$ is a complex conjugate of $E_\theta$. Then
using the following relation \be \int\limits_{-\infty }^{\infty
}\exp \left\{ i\left( k_{z}-k_{z}^{\prime }\right) z+i\left(
s-s^{\prime }\right) \theta \right\} d\theta dz=\left( 2\pi
\right) ^{2}\delta \left( k_{z}-k_{z}^{\prime }\right) \delta
\left( s-s^{\prime }\right), \label{deltaf1} \ee we obtain the
emissivity of the charged particle in the form \be
\frac{d^{3}P}{dtd\omega dk_{z}}=-\frac{1}{\left( 2\pi \right)
^{4}}\int\limits_{-\infty }^{\infty }rdrJ_{\theta
}^{\rm{part}}\left( \omega ^{\prime },r,s,k\right) E^*_{\theta
}\left( \omega ,r,s,k\right) e^{-i\left( \omega -\omega ^{\prime
}\right) t}dsd\omega ^{\prime }. \label{emis1} \ee Now we can
integrate over $s$ and $\omega ^{\prime }$ using \be
\int\limits_{-\infty }^{\infty }\delta \left( \omega ^{\prime
}-s\frac{V}{R}\right) \delta \left( s-\omega \frac{R}{V}\right)
e^{-i\left( \omega -\omega ^{\prime }\right) t}dsd\omega ^{\prime
}=1, \label{deltaf2} \ee and over $r$ using delta-function
$\delta(x-X)$ in the expression for $J^{\rm part}_{\theta}$ (see
eq.~[\ref{FJpart}]). Then substituting $\omega$ and $k_z$ by the
dimensionless values $\nu$ and $\mu$ (eq.~[\ref{numu}]) we obtain
equation (\ref{eta}).

{}
\newpage
\newpage
\newpage
\end{document}